\documentclass[aps,twocolumn]{revtex4}

\usepackage{epsfig}

\begin{document}

\title{Robustness of first-order phase transitions in one-dimensional 
long-range  contact processes}

\author{Carlos E. Fiore and M\'ario J. de Oliveira} 

\affiliation{Departamento de F\'{\i}sica, Universidade Federal do Paran\'a \\
Caixa Postal 19044, 81531-000 Curitiba, Paran\'a, Brazil}

\affiliation{Instituto de F\'{\i}sica, Universidade de S\~ao Paulo \\
Caixa Postal 66318,
05315-970 S\~{a}o Paulo, Brazil}

\date{\today}

\begin{abstract}
It has been proposed (Phys. Rev. E {\bf 71}, 026121 (2005)) 
that unlike the short range contact process, a long-range 
counterpart may lead to the existence a discontinuous
phase transition in one dimension. 
Aiming at exploring such link, here  we investigate thoroughly a
family of   long-range contact processes.
They are introduced through
the transition rate $1+a\ell^{-\sigma}$, where  $\ell$ is the length
of   inactive islands surrounding particles. 
In the former approach we  reconsider  the original 
model (called $\sigma-$contact 
process),  by considering distinct mechanisms of  weakening   
the long-range interaction toward the short-range limit. 
 Second, we study  the effect of different  rules, including 
creation and annihilation by clusters of particles and  distinct
versions with infinitely many absorbing states.
Our results show that  all examples presenting a single absorbing state, 
 a discontinuous transition is possible for small $\sigma$. On the other hand,
the presence of infinite absorbing states leads to distinct scenario depending
on the interactions at the frontier of inactive sites.


\end{abstract}

\maketitle

\section{Introduction}
Nonequilibrium  phase transitions into absorbing states 
 describe a large sort of phenomena including
 chemical reactions, disease spreading, competition between
species, wetting processes, calcium dynamics and others \cite{marro,munoz4}.
 Due the absence of equilibrium analogous, 
they  are essential in the  establishment of  the 
main ingredients required for the emergence of phase transitions 
and critical behavior.

The contact process (CP) \cite{harris} is probably the best
example  yielding an absorbing phase transition. 
Particles are  created catalytically, but are spontaneously annihilated. 
It presents a set of critical exponents
belonging to the directed percolation (DP) universality
class. The DP conjecture \cite{janssen} embraces not
only the CP, but also generic absorbing phase
transitions with no extra symmetries and conservation laws. Such 
examples are reaction-diffusion processes, cellular-automata models 
and even continuous descriptions with multiplicative noise 
\cite{mn,odor04}. 
More recently it has been observed experimentally in turbulent nematic
liquid crystals \cite{sano}.  

Differently from the continuous case, 
one-dimensional discontinuous absorbing transitions 
  has been much less  observed. Except in special
cases \cite{lip00,lee09,fibjp}, 
its manifestation in short range systems 
have been the subject of a longstanding controversy 
\cite{dic91,fiore04,card06,hinri00,park08}. 
The absence of a discontinuous transition 
would stem from  the suppression of compact clusters
coming from the  large fluctuations in one dimension. On the 
other hand, by increasing the dimensionality,
the fluctuations are less relevant and
the formation of compact clusters becomes possible.

Long-range  interactions have  been proposed
like more realistic descriptions in different nonequilibrium
phenomena, when compared 
with their short-range counterparts. Some examples
of systems  presenting
long-range interactions include wetting phenomena,
spreading of diseases over long distances and others 
\cite{how01,vern03,mollison}.
As an effect of long-range interactions, the absorbing transition
may deviate from the original DP case and
belong to the different  universality classes \cite{grass86,jan99,hinri99,
tessone,gine06,hinri07}.  
Other  remarkable difference concerns 
 the possibility of stabilizing compact clusters in one dimension.
Ginelli et al. \cite{gine05} introduced the $\sigma-$contact process,
in which  particles are  created and annihilated like
the usual short-range CP, but
the activation rate depends the length $\ell$ 
of the island of inactive sites  according to expression
$1+a\ell^{-\sigma}$.
They found that  for   $0<\sigma<1$ 
the interactions are effectively long-range and  
the phase transition becomes discontinuous. On
the other hand, for $\sigma>1$  the long-range parameter play 
not relevant role
and the phase transition remains second-order (similar to  its 
short-range version).

Despite the  study of  the $\sigma-$CP 
under different methodologies \cite{gine05,fiore07}
some aspects have not been addressed so far. 
Is it present a first-order
transition   in the limit of extreme weak ($a<<1$)
long-range interactions? Does the competition with short
range interactions become the system  still able to suppress fluctuations
that destabilize compact clusters? 
Is the phase coexistence 
 maintained by  changing the interaction rules?
Does the existence of infinitely absorbing states influence 
the order of transition?

Aiming to answer above queries, in this paper we investigate thoroughly a
family of one-dimensional long-range contact models.
First, we
reconsider the $\sigma-$CP by  weakening sufficiently the  
long-range interaction toward 
the short-range limit and further by introducing  an effective 
competition with   short-range interactions.  Although 
the emergence of a discontinuous transition 
is  expected not depending on the parameter $a$ \cite{gine05}, 
a quite interesting point  
would concern the stabilization 
of compact clusters over  extreme  small long-range interactions 
(thus close to the short-range regime).  
In such cases the long-range should act as a small (but relevant) perturbation.
Second,  we consider the effect of  different   interaction rules, including
creation and annihilation in the presence of clusters of particles
(instead of one particle case) and  infinitely many absorbing states. 
These models are long-range versions of the named pair-creation CP, 
pair-annihilation (PAM), triplet-annihilation models (TAM)
and the pair contact process (PCP), respectively \cite{dic89,dic91,jensen}.
All models will be studied over mean-field approximations and 
extensive numerical
simulations in the constant rate (ordinary) \cite{marro} 
and the constant particle number (conserved) 
ensembles \cite{tome01,hilh02,oliv03,fiore05,fiore07}. 
Our results show that for all systems with a single
absorbing state the occurrence of
discontinuous transitions is held by decreasing $\sigma$. 
For the long-range PCP, on the other hand, different scenario
are possible. By measuring the length $\ell$ between extreme
pairs of particles, the transition becomes first-order for
low $\sigma$. On the other hand, when $\ell$  is
the distance between a pair and the nearest
particle surrounding the inactive island, 
the transition is always continuous. 

This paper is organized as follows: In Sec. II we described 
all  methods, in Sec. III we present the models
and numerical results and finally conclusions are showed
in Sec. IV.
\section{Constant rate and conserved ensembles}
The one-dimensional  
contact process  is defined in a chain of $L$ sites 
where each site $i$ is attached by a two-state 
occupation variable $\eta_i$ reading $\eta_i=0$ or $1$, according
to whether it is empty or occupied, respectively. 
Interaction rules are composed of creation
and annihilation of   particles, represented by 
transition rates $\omega_i^c$ and $\omega_i^a$. 
Particles are created in empty active sites and are spontaneously
annihilated.

Systems will be studied 
in the constant rate (ordinary) and in the constant particle number
(conserved) ensembles 
\cite{tome01,oliv03,fiore05,fiore07}. In the former case
the control parameter (the creation 
or annihilation rates) are held fixed but
the total particle number ${\bar n}$ fluctuates. It is described by 
the total transition rate $w_i$ 
\begin{equation}                                                               
w_i= \omega_i^c+\alpha \omega_i^a,                         
\label{eq1}                                                                   
\end{equation}
where $\alpha$ denotes the strength of the annihilation rate.
For low $\alpha$,  phase is active, in which
 particles are continuously created and annihilated. By increasing 
$\alpha$, a phase transition into an absorbing state takes place.
Except the PCP, the absorbing state
is characterized by a full empty lattice. For the PCP, any configuration
devoid of pairs is absorbing.

The transition point and the nature of transition
can be precisely  identified by performing 
spreading simulations \cite{marro}. Starting from an initial seed, it  
consists of determining  the time evolution of appropriate quantities, such
as the survival probability $P_s(t)$, the total number of particles
$N(t)$ and the mean square spreading $R^2(t)$.

At the emergence of phase transition, these quantities 
follow algebraic behaviors given by
\begin{equation}                                                               
P_s(t)\sim t^{-\delta}, \quad N(t)\sim t^{\eta} \quad  {\rm and} \quad     
R^2(t)\sim t^{2/z},                                                              
\label{eq1} 
\end{equation}
where $\delta,\eta$ and $z$ are their associated dynamic critical
exponents. 
For second-order DP phase transitions, they
 read $\delta=0.159464(6)$, $\eta=0.313686(8)$ and 
$z=1.580745(10)$ \cite{odor04}.
In a discontinuous transition,  despite the order
parameter gap, they also present algebraic behaviors with
critical exponents  given by $\delta=1/2$, $\eta=0$ and $z=1$, 
which is compatible with the Glauber-Ising (GI) model
 \cite{odor04}. 

Thus, spreading simulations provides  not only locating 
the transition point, but also
to classify the order of transition. The change in the order
of transition will be  characterized by an alteration 
in the critical exponents.
Off the transition regime, above dynamic quantities deviate from the
power law behaviors.

For systems with infinitely many absorbing states \cite{dic93}, spreading
experiments become particularly hard to be used. In particular,
the dynamic exponents  $\delta,\eta$ and $z$ 
are 
strongly dependent on the initial conditions, presenting non-universal
values 
\cite{dic93,munoz1,munoz3,munoz2}.  A simpler procedure for 
locating the critical point in such cases is to
 study the  order-parameter  decay starting from
a fully occupied initial condition. Unlike
above exponents, $\theta$ does not depend on the initial
configuration. One 
expects $\phi$ behaves as $\phi \sim t^{-\theta}$ 
at the critical point, where $\theta=\delta$. Conversely, 
we should expect a
non power-law behavior at a discontinuous transition. 
Since alternative methods (e.g hysteretic ones) 
can not be used for systems with absorbing states,   
another strategy is required. In particular, 
we have calculated the probability distribution 
(in the steady regime) by considering different initial
configurations. A bimodal distribution reveals the discontinuous
transition.

In the constant rate ensemble,   
the control parameter is the total particle number $n$.
Particles  are created as in the ordinary case, but 
instead of creating new particles, a cluster of $k$ sites leaves
its place and jumps to $k$ active sites.  
One may define the conserved ensemble as a $2k$-site process, 
in which   creation and annihilation occurs simultaneously 
according to the following transition  \cite{fiore05}
\begin{equation}
w = 
\omega_i^a\underbrace
{\omega_j^c\omega_l^c...\omega_m^c}_{k-creation \quad processes}.
\label{eq4}
\end{equation}
It has been shown \cite{oliv03,fiore05} that  
in the thermodynamic limit above
dynamics is equivalent to that studied in the  constant rate ensemble. 
The parameter ${\bar\alpha}$ fluctuates
and is calculated through expression
\begin{equation}
\label{eq5}
{\bar\alpha} = \frac{\langle\omega_j^c\rangle_c}
{k\langle\omega_i^a\rangle_c},
\end{equation}
where $\langle...\rangle_c$ denotes a generic average evaluated
over the conserved ensemble.
An immediate advantage  concerns
its simplicity for locating the transition point.
In this case, by considering the system
constrained in the subcritical regime, e. g, a finite number of particles
placed in an infinite lattice, the addition of particles drives  the
system toward the transition point $\alpha_0$ according to the expression
 \cite{brok99,tome01,fiore05,fiore07}, 
\begin{equation}
{\bar \alpha}-\alpha_0 \sim n^{-1}.
\label{eq55}
\end{equation}
Thus, from the above,  the transition point is obtained
by a linear extrapolation in $1/n$.
Other advantage refers the classification of
the  phase transition,  obtained by  measuring
 the particle structures for different $n$.
For second-order transitions,
the clusters are fractals \cite{vics92}, whereas they become compact  in
a discontinuous transition. 
 Being $R$ the mean distance  of particles located at extremities
of the system, we have that \cite{marro,fiore05,fiore07} 
\begin{equation}
R \sim n^{1/d_F},
\label{fractal}
\end{equation}
where $d_F$ is the fractal dimension.
 For one-dimensional
systems belonging to the DP universality
class $d_F$ reads $0.74792...$, whereas  at the phase coexistence
it is the proper euclidean dimension $d=1$
(consistent to the existence of   compact clusters). 
The above values are related with  dynamic exponents
through expression $d_{F}=2(\eta+\delta)/z$. 



\section{mean field approximation}
\begin{figure}
\centering
\epsfig{file=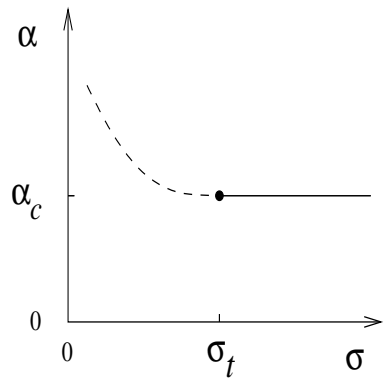,width=5cm,height=5cm}
\caption{The pair mean-field approximation phase diagram
in the space $\alpha$ versus $\sigma$.
Continuous and dashed lines denote second and first order
phase transitions, respectively 
and the full circle denotes the tricritical point.}
\label{dfmf}
\end{figure}

Before performing simulations we have 
analyzed the models by means of a cluster
approximation at the level of two nearest 
neighbor sites \cite{fiore07}. In this case, the system is
described by the one-site probabilities $P(0)$
and $P(1)$ and the two-site probabilities
$P(11)$, $P(10)$, $P(01)$ and $P(11)$. However,
only two of them are independent. 
The   generic probability of a string
of $\ell$ consecutive sites is approximated by
\begin{equation}
P(\eta_1,\eta_2,\eta_3,\ldots,\eta_\ell) =
\frac{P(\eta_1,\eta_2)P(\eta_2,\eta_3)\ldots 
P(\eta_{\ell-1},\eta_\ell)}
{P(\eta_2)P(\eta_3)\ldots P(\eta_{\ell-1})}.
\end{equation}
By choosing $P(1)=\langle\eta_i\rangle$ 
and $P(11)=\langle\eta_i\eta_{i+1}\rangle$ and taking 
account the translation invariance,   their  evolution
equations  read
\begin{equation}
\frac{d}{dt}\langle\eta_i\rangle =
\langle(\bar{\eta}_i-\eta_i)w_i(\eta)\rangle,
\label{pair}
\end{equation}
and
\begin{equation}
\frac{d}{dt}\langle\eta_i\eta_{i+1}\rangle =
\langle(\bar{\eta}_i-\eta_i)\eta_{i+1}w_i(\eta)\rangle +
\langle\eta_i(\bar{\eta}_{i+1}-\eta_{i+1})w_{i+1}(\eta)\rangle,
\end{equation}
where $\bar{\eta}_i \equiv 1-\eta_i$.
 By using the approximation (\ref{pair}), we get a set
of two closed equations for $P(1)=\rho$ and $P(11)=\phi$.
At the stationary state we found for all models
 studied a general structure in the relation for $\alpha$ vs $\rho$
given by 
\begin{equation}
\alpha\rho = \alpha_c\rho + A \rho^2 
\end{equation}
up to order $\rho^2$, where $\alpha_c$ is a numerical constant
and $A$ depends on the other parameters but not on $\alpha$.
From this equation it follows that a critical line  
occurs at $\alpha=\alpha_c$ and $A<0$. When $A>0$, the transition
becomes first-order and a tricritical point occurs at
$A=0$. The phase diagram is of the type shown in Fig. \ref{dfmf}.

\section{Numerical results}
Except for the PCP model, numerical simulations will be performed
for large system sizes ($L=2^{16}$) and periodic boundary conditions.
   In the conserved case,  
 MC simulations are started  by constraining
the system in the subcritical (absorbing) regime. In practice, it is
done by taking finite $n$'s in a large $L$ and  check whether a particle
touches the border. If a particle reaches the border,
we increase $L$.
By simulating distinct $n$'s (with ${\bar \alpha}$ 
 computed from Eq. (\ref{eq55})),  the transition point  $\alpha_0$ 
is obtained by means of a linear extrapolation in
$1/n$. The  nature of phase transition is  identified by 
calculating the fractal dimension,  measured 
from the dependence of  $R$ on $n$. Further, we 
check above results by performing   epidemic simulations starting
from an initial seed  in which  the transition point $\alpha_0$ is  
located by identifying algebraic behaviors for  $N$ and 
$P_s$. Their corresponding dynamic exponents $\eta$ and $\delta$ 
classifies the order of transition.

\subsection{Long-range contact process ($\sigma-$CP)}
In the  usual CP,  particles are created
in empty sites surrounded by at least one particle  and 
are spontaneously annihilated. It is defined by the transition rates
\begin{equation}                                                               
\omega_i^c=\frac{1}{2}(1-\eta_i)\sum_{\delta}\eta_{i+\delta},
\label{rate1}
\end{equation}
and
\begin{equation}
\omega_i^a=\eta_{i},
\label{rate2}
\end{equation}
 for  particle creation and annihilation, respectively.
In the $\sigma-$CP the creation rate is replaced by the following
expression
\[
\omega_i^c
=\frac12\, 
\sum_{\ell=1}^\infty  (1+\frac{a}{\ell^\sigma})\eta_{i-1} {\bar\eta}_i
{\bar\eta}_{i+1}\ldots{\bar\eta}_{i+\ell-1}\eta_{i+\ell}
\]
\begin{equation}
+\frac12\,  
\sum_{\ell=1}^\infty (1+\frac{a}{\ell^\sigma})\eta_{i+1}{\bar\eta}_i 
{\bar\eta}_{i-1}\ldots{\bar\eta}_{i-\ell+1}\eta_{i-\ell},
\label{rate3}
\end{equation} 
which depends on  $\ell$, $a$ and $\sigma$ 
 and ${\bar\eta}_i\equiv 1-\eta_i$.
For $a=0$, one recovers   the original short-range CP, whose 
second-order phase transition occurs at $\alpha_c=0.303227...$ \cite{marro}.  

\begin{figure}
\centering
\epsfig{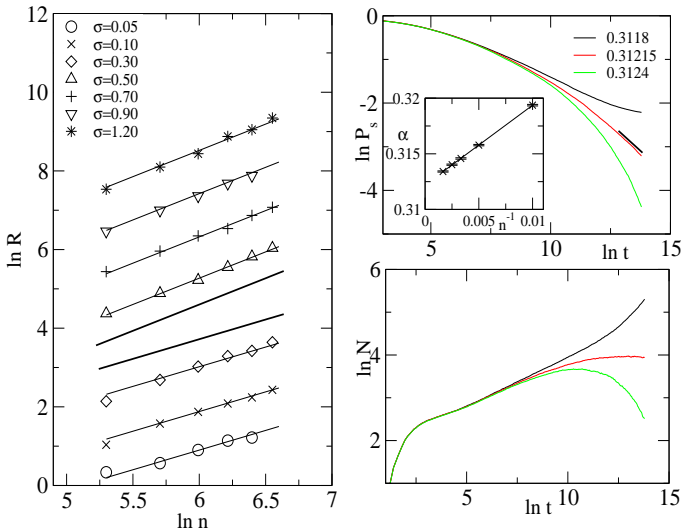}
\caption{In the left, log-log plot of the average cluster size $R$ versus
the total number of particles $n$ for several values of $\sigma$
for $a=0.05$. The up and down predicted curves 
have slopes 1.33704 and 1, respectively. The data points
have been shifted in order to avoid  overlapping. 
In the right, log-log plot of $P_s$ and $N$
for distinct values of $\alpha$ for $\sigma=0.1$. The predicted 
asymptotic slopes  are consistent with GI values. In the
inset parameter ${\bar \alpha}$ versus $n^{-1}$ in the conserved ensemble.}
\label{fig0}
\end{figure}

Here we  have  weakened the long-range
interaction toward the short-range limit, in order to see if
a phase coexistence  still exists for  $a<<1$.
In Fig. \ref{fig0}, we show  results  for $a=0.05$.
Note that  the fractal dimension changes (from $0.75$ to $1$)  
for  $\sigma < \sigma_{t}=0.4(1)$,
consistent with the emergence of 
a discontinuous transition for smaller $\sigma$.
Spreading experiments,
 showed in the right part for $\sigma=0.1$, confirms
such conclusion.  At  $\alpha_0=0.31215(5)$,
 both quantities $P_s$ and  $N$ present algebraic behaviors consistent 
 with exponents $\delta=1/2$ and $\eta=0$, respectively. 
The above estimate  agrees very well with the conserved 
ensemble result $\alpha_0=0.31221(3)$.
In addition, we have also studied the possibility of discontinuous transitions
 for  more extreme cases. For $a=0.01$ and $\sigma=0.005$, the  transition
is first-order, yielding at 
$\alpha_0=0.30572(3)$, a value rather close to the short-range case $0.303227$.
 For completeness, we have 
considered the opposite case, e.g. the occurrence
of discontinuous transition for  larger $a$'s. 
Our results (not shown) support that the first-order transition line
moves toward larger $\sigma$'s. For example, for $a=5$ and $\sigma=1.2$,
the transition is first-order yielding at $\alpha_c=0.439655(5)$.
The crossover  occurs at $\sigma_t=1.3(1)$, which is
larger than $\sigma_t=1.0(1)$ for $a=2$.

Above conclusions are also predicted by pair mean-field results.
It gives a critical line at the value $\alpha_c=1/2$
and    the tricritical point occurring at
\begin{equation}
\zeta(\sigma_t) = \frac{1+a}{a},
\end{equation}
where $\zeta(\sigma)$ is the Riemann zeta function defined by
$\zeta(\sigma)=\sum_{k=1}^{\infty}k^{-\sigma}$.
From the above, it follows that $\sigma_t>1$, in
accordance with numerical results for larger $a$ but not
for sufficient small $a$'s.

\begin{figure}
\centering
\epsfig{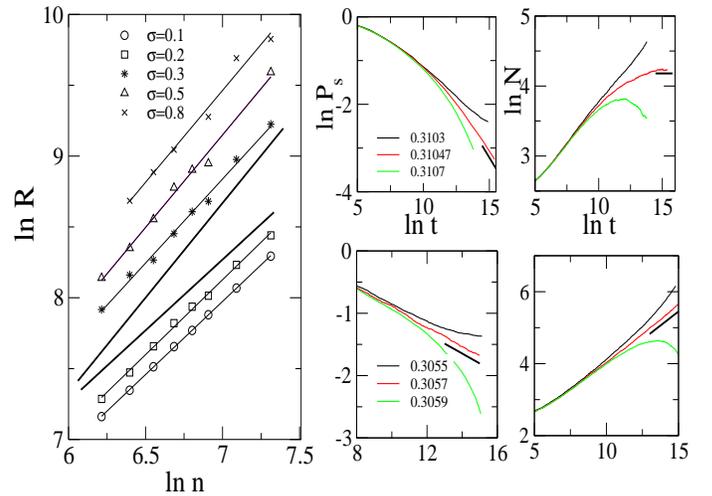}
\caption{In the left, log-log plot of the average cluster size $R$ versus
the total number of particles $n$ for several values of $\sigma$
at the subcritical regime. The up and down straight lines 
have slopes 1.33704 and 1, respectively. The data points
have been shifted in order to avoid  overlapping. 
In the right, from top to bottom, log-log plot of the time evolution
of $P_s$
and $N$ for distinct values of $\alpha$ 
for $\sigma=0.1$ (top)  and $\sigma=0.3$ (bottom), respectively. 
The predicted asymptotic slopes 
 are consistent with GI (top) and DP (bottom) values.}
\label{fig2}
\end{figure}

\begin{figure}
\centering
\epsfig{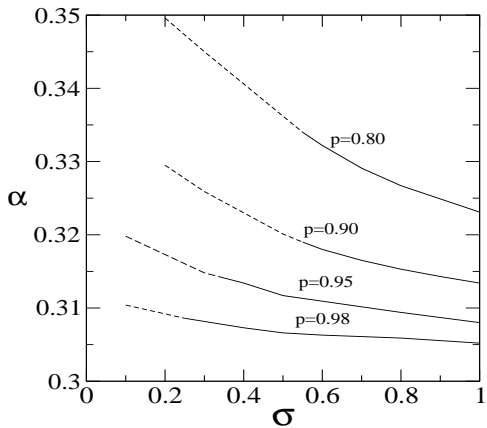}
\caption{Parameter $\alpha$ versus $\sigma$ for different values
of $p$. Continuous and dashed lines denote second and first-order
transitions, respectively. For each $p$, absorbing and active 
phases are located
above and bellow of corresponding lines, respectively.}
\label{fig1}
\end{figure}

Further, we introduce the competition 
with short-range interactions. This will be accomplished by
performing short and long-range
processes  probabilities  $p$ and  $1-p$, respectively. 
For low $p$, one expects a qualitative  behavior 
similar to the full long-range case, whereas for extreme large $p$
if a change in the transition occurs, it should manifest for 
sufficient low $\sigma$. 
In Fig. \ref{fig2} we examine the  phase transitions for $p=0.98$
and different $\sigma$.
As in the previous case, the system structure also 
changes by decreasing $\sigma$
and  clusters  become compact for  $\sigma <0.3$.  
This is also checked by comparing the time evolution of $P_s$ and $N$ 
for $\sigma=0.3$ and $\sigma=0.1$.  At the
transition points $\alpha_c=0.3057(1)$ and $\alpha_0=0.31045(5)$, 
above quantities 
present distinct algebraic behaviors  consistent
with the DP and GI exponents.
In Fig. \ref{fig1} we show  the phase diagram 
for distinct (but large) values of $p$. 
As expected, the  coexistence line move toward lower values of $\sigma$.

In summary, above ``weakening''  approaches 
show that it suffices a small long-range ``perturbation'' in the original
CP to provoke a change in the order of transition.
\subsection{Long-range pair-creation, pair and triplet 
annihilation contact models}
\begin{figure}
\centering
\epsfig{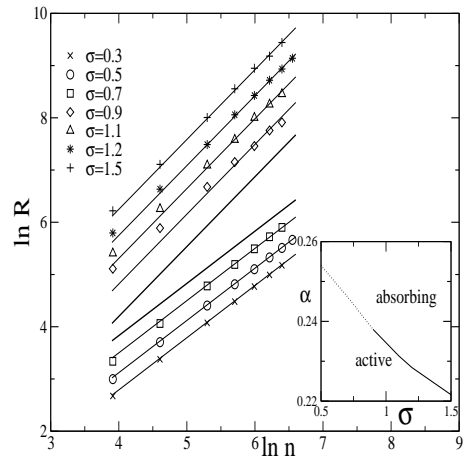}                                
\caption{Log-log plot of $R$ versus $n$ for                                    
the long-range pair-creation CP for
 distinct values of $\sigma$.   The up and down straight lines 
have slopes 1.33704 and 1, respectively and they              
have been shifted in order to avoid overlapping. In the
inset, we show its phase diagram. Continuous and dotted lines
denote continuous and discontinuous transition, respectively.}
\label{fig4-1}
\end{figure}

Here we study the effect of distinct interaction
rules in the long-range CP.
The first change, called  $\sigma-$pair CP, is similar 
to the $\sigma-$CP  but new 
particles can be created
 only in empty sites surrounded by pairs of particles. 
The creation rate $\omega_{i}^{c}$ reads
\[
\omega_i^c 
=\frac12\, 
\sum_{\ell=1}^\infty  (1+\frac{a}{\ell^\sigma})\eta_{i-2}\eta_{i-1} {\bar\eta}_i
{\bar\eta}_{i+1}\ldots{\bar\eta}_{i+\ell-1}\eta_{i+\ell}
\]
\begin{equation}
+\frac12\,  
\sum_{\ell=1}^\infty (1+\frac{a}{\ell^\sigma})\eta_{i+2}\eta_{i+1}{\bar\eta}_i 
{\bar\eta}_{i-1}\ldots{\bar\eta}_{i-\ell+1}\eta_{i-\ell}.
\label{rate6}
\end{equation} 
The limit $a=0$ corresponds to the short-range case, 
in which a  DP phase transition  
yields at $\alpha_c=0.13397(4)$ \cite{fiore05}. In 
Fig. \ref{fig4-1}, we show the main  results for distinct
$\sigma$'s and $a=2$.
The transition is also continuous  for $\sigma>1$ and
becomes first-order for  $0<\sigma<1$. However, as an 
effect of the creation in the presence of pairs, the cluster are somewhat 
more compact than the usual $\sigma-$CP. For example, for
 $\sigma=0.5$ the cluster density $\rho=N/R$ (evaluated
from the inverse of slopes of curves $R$ vs $n$)
is about $0.78$ for the $\sigma-$CP, whereas it reads $0.81$ 
for the $\sigma-$pair CP. However, the creation by pairs of particles
is not sufficient effective for shifting the coexistence line for larger
$\sigma$. In contrast with the
$\sigma-$CP, 
above results are not predicted by  pair mean-field approximation,
in which  the phase transition is always first order. On
the other hand, when $\sigma\to\infty$
the parameter $\alpha\to 1/4$, which is in accordance with the mean-field
short-range case.

Next, we consider the opposite situation, in which 
 particles are created like the above
$\sigma-$CP, but only pairs of particles are allowed to be  annihilated. 
The annihilation rate  $\omega_i^a$ reads  
\begin{equation}
\omega_i^a=\eta_{i}\eta_{i+1}.
\label{rate4}
\end{equation}
In the conserved ensemble,
Eq. (\ref{eq5}) is used for calculating for ${\bar \alpha}$
by taking $k=2$.
For $a=0$ one recovers the original
pair-annihilation contact model (PAM), in which a 
 DP continuous phase transition
yields  at $\alpha_c=0.18622(3)$ \cite{dic89,fiore05}.

\begin{figure}
\centering
\epsfig{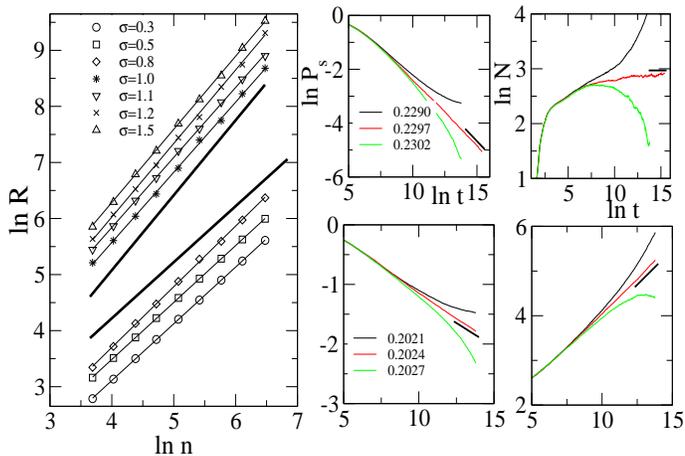}                               
\caption{Log-log plot of $R$ versus $n$ for                                    
the long-range PAM and distinct values of $\sigma$.                            
Straight lines have slopes $1.33704$ (up) and $1$ (down), 
respectively and they              
have been shifted in order to avoid overlapping.
In the right, from top to bottom, log-log plot of the time evolution           
of $P_s$ and $N$ for distinct  $\alpha$'s                 
for $\sigma=0.3$ (top)  and $\sigma=1.2$ (bottom), respectively.     
The predicted asymptotic slopes 
 are consistent with GI (top) and DP (bottom) values.}
\label{fig4}
\end{figure}

Like the $\sigma-$CP, for $a=2$ the phase transition
becomes first-order for $0<\sigma<1$. The crossover between continuous
and discontinuous occurs between 0.8 and 1.1.
In Fig. \ref{fig4} we show the main results for distinct $\sigma$'s. 
As an consequence of the
pair annihilation, the compact clusters are less dense than the previous
cases (e. g., for $\sigma=0.5$ the cluster density $\rho$ is about $2/3$).

\begin{figure}
\centering
\includegraphics[scale=0.35]{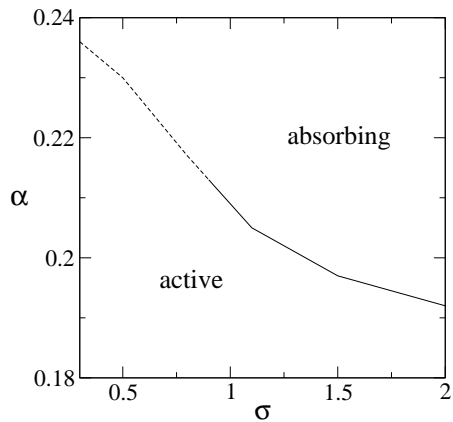}
\caption{Parameter $\alpha$ versus $\sigma$ for 
the long-range PAM. Continuous and dashed lines denote second and first-order
transitions, respectively.}
\label{fig3}
\end{figure}

Using Eq. (\ref{eq55})  
we have built the phase diagram shown in Fig. \ref{fig3}. 
Since $\langle\omega_i^c\rangle_c$
 is proportional to  $\ell^{-\sigma}$, it increases by decreasing 
 $\sigma$. On the other hand, the average 
 $\langle\omega_i^a\rangle_c$ also increases, as a result
of  more compact particle displacements. The
competition between averages  results in a net increase of $\alpha$
with the decrease of $\sigma$. 
In similarity to the $\sigma-$CP 
the pair mean-field approximation also gives
first and second order transitions, with the tricritical point given
by
\begin{equation}
\zeta(\sigma_t) = 4\frac{1+a}{a}.
\end{equation}
Again, from this equation it follows that $\sigma_t>1$. 
The pair mean-field predicts a critical line
at  $\alpha_c=1/2$.

\begin{figure}
\centering
\epsfig{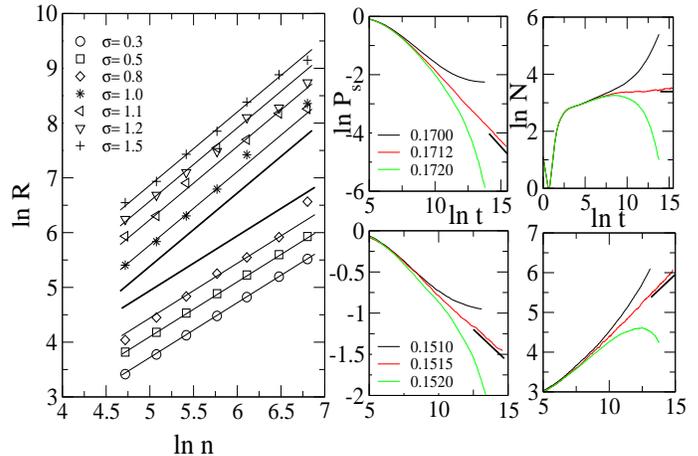}
\caption{Log-log plot of $R$ versus $n$ for 
the long-range TAM and distinct values of $\sigma$.
Straight lines have slopes $1.337$ (up) and $1$ (down), 
respectively and they have been shifted in order to avoid overlapping.
In the right, from top to bottom, log-log plot of the time evolution           
of $P_s$ and $N$ for distinct values of $\alpha$                  
for $\sigma=0.3$ (top)  and $\sigma=1.2$ (bottom), respectively.               
The predicted asymptotic slopes in the top                  
and bottom are consistent with GI and DP values.}
\label{fig5}
\end{figure}

\begin{figure}
\centering
\includegraphics[scale=0.35]{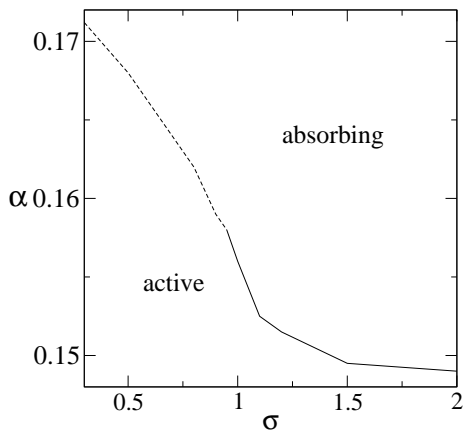}
\caption{Parameter $\alpha$ versus $\sigma$ for 
the long-range TAM. Continuous and dashed lines denote second and first-order
transitions, respectively.}
\label{fig6}
\end{figure}

Now we consider the influence of annihilation of 
three adjacent particles. This study is motivated
by previous works \cite{dic89,fiore05}  which shows that the inclusion
of triplet annihilation brings great differences in the phase
diagram, when compared with single and pair annihilations.
The transition rate $\omega_i^a$ is then given by
\begin{equation}
\omega_i^a=\eta_{i-1}\eta_{i}\eta_{i+1},
\end{equation}
and particles are created like the $\sigma-$CP.
In the conserved ensemble, 
Eq. (\ref{eq5}) is used for calculating for ${\bar \alpha}$ 
for $k=3$.
The short-range case ($a=0$) has been extensively studied at 
Refs. \cite{dic89,fiore05}, where
a continuous phase transition belonging
to the DP universality class takes place at $\alpha_c=0.14898(5)$.
In Figs. \ref{fig5} and \ref{fig6}, we plot the average 
cluster size $R$ versus $n$ and the phase diagram
for different values of $\sigma$. As in previous cases,
 the phase transition  is second-order for $\sigma>1$,
becoming first-order for $0<\sigma<1$. The crossover between continuous
and discontinuous takes place between 0.8 and 1.1. 
As an effect of the triplet annihilation, 
the compact clusters are less dense than all
previous cases (for $a=2$).

Using the same procedure adopted previously,
we have built the phase diagram shown in Fig. \ref{fig3}. 
The transition points $\alpha$ varies mildly with $\sigma$, as an
effect of  simultaneous increase of $\langle\omega_i^c\rangle_c$ and 
 $\langle\omega_i^a\rangle_c$   by decreasing $\sigma$. 
For smaller $a$, the system also presents a phase coexistence, although
the crossover seems occur for smaller $\sigma$, when compared with 
the case $a=2$.
For example, for $a=0.1$ and $\sigma=0.1$, the phase coexistence
yields at $\alpha_0=0.1558(1)$.

As in the previous cases, the pair mean-field approximation reproduces
first- and second-order transitions, with a critical line and
 tricritical point occurring  at $\alpha_c=2/3$ and
\begin{equation}
\zeta(\sigma_t) = 9\frac{1+a}{a},
\end{equation}
respectively. Also, as in previous cases, it follows that $\sigma_t>1$.
In summary, the existence of a discontinuous transition
in both PAM and TAM  for $0<\sigma<1$ indicate 
that  the pair and triplet annihilations does not provoke
sufficient strong fluctuations that would lead the suppression
of compact clusters.
\subsection{Long-range pair contact process ($\sigma-$PCP)}
In the pair contact process (PCP), 
only pairs of particles can be annihilated or creating new  particles. Unlike
all previous models, any configuration
devoid of pairs is absorbing and thus the PCP displays infinitely
many absorbing states. The order parameter is 
 pair density $\phi$ instead of the particle density $\rho$. 
The PCP model has been extensively studied in the last
years \cite{jensen,fiore05} and despite the differences
with the all previous models, its  absorbing   
transition belongs to the DP universality class.  
Being $p$ the probability of annihilating pairs of particles,
the phase transition yields at
at $p_c=0.077090(5)$  \cite{jensen,fiore05}.
The parameters $p$ and $\alpha$ (used here) are related through expression
$p=\frac{\alpha}{\alpha+1}$.

The long range version can be introduced similarly than all previous cases. 
However, in order to investigate the role  of infinitely absorbing
states, we take  two different  cases. In the former, 
the activation rate is given by Eq. (\ref{rate6}), 
implying that the distance $\ell$ is measured up to the nearest
particle at the edge of an inactive island.
The latter  takes into account the distance up to the nearest 
pair given by
\[
\omega_i^c 
=\frac12\, 
\sum_{\ell=1}^\infty  (1+\frac{a}{\ell^\sigma})\eta_{i-2}\eta_{i-1} {\bar\eta}_i
\ldots\eta_{i+\ell}\eta_{i+\ell+1}
\]
\begin{equation}
+\frac12\,  
\sum_{\ell=1}^\infty (1+\frac{a}{\ell^\sigma})\eta_{i+2}\eta_{i+1}{\bar\eta}_i 
\ldots \eta_{i-\ell-1}\eta_{i-\ell}.
\label{rate7}
\end{equation} 
In both cases, the annihilation rate is given by Eq. (\ref{rate4}).

\begin{figure}
\centering
\epsfig{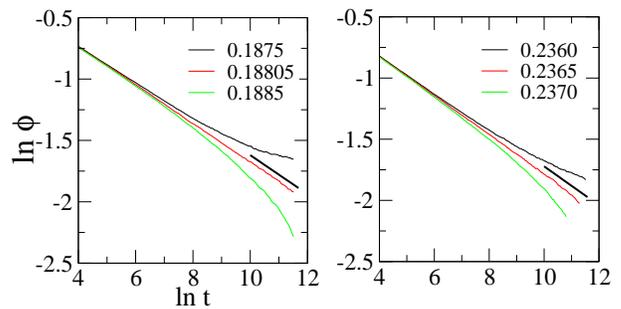}
\caption{From the left to right, log-log plot of 
the time evolution of pair density $\phi$ 
starting from a fully occupied lattice for  $\sigma=0.5$ and $\sigma=0.1$
for different $\alpha$'s. The straight
lines in the middle curves have slopes consistent with $0.159464(6)$. }
\label{fig7}
\end{figure}

\begin{figure}
\centering
\epsfig{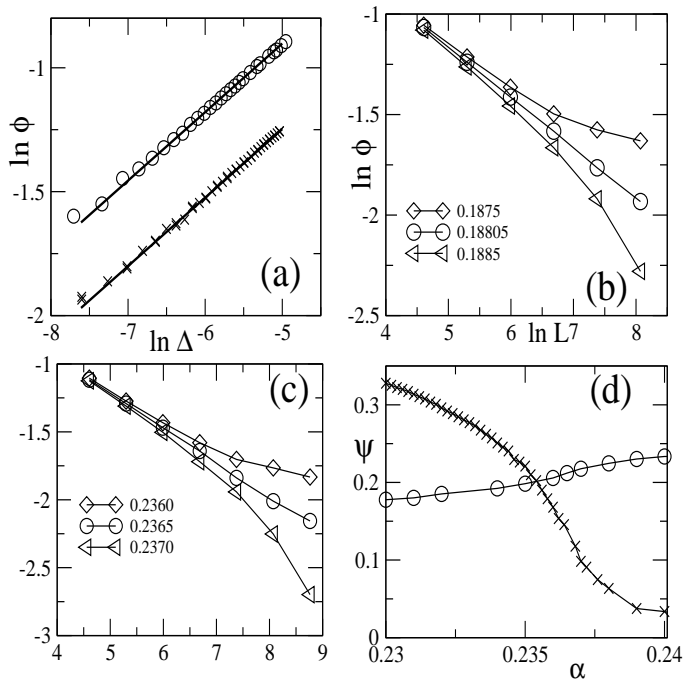}
\caption{In $(a)$, the log-log plot of $\phi$
versus $\Delta \equiv \alpha_c-\alpha$ 
for different $\sigma$ and $L=3200$. The straight
lines have slope $0.276486$. Log-log plot
of $\phi$ versus $L$ for  $\sigma=0.5$ $(b)$ and  $\sigma=0.1$ $(c)$.
The straight lines have slopes $0.2520718$. In $(d)$ we plot
the dependence of
$\psi$ on $\alpha$, where $\psi=\phi$ (stars) and $\psi=\rho-\phi$
(circles)  for $\sigma=0.1$.  }
\label{fig8}
\end{figure}

Since the existence of infinite absorbing states makes
the use of spreading simulations difficult,
we  adopt  the procedure described in Sec. II,
consisting of studying   the time evolution of the  pair density
$\phi$ starting from a fully occupied lattice. In Fig. \ref{fig7} 
we plot the decay of $\phi$ for different $\sigma$'s by
taking the first version. 
We  focus the analysis for $a=2$ and low $\sigma$, in
order to see the effect of strong  long-range interactions.
We see that for both cases,  the decay of $\phi$ is algebraic 
  at $0.18805(5)$ ($\sigma=0.5$) and $0.2365(1)$ ($\sigma=0.1$)
with   exponents   consistent with the DP value $\theta=0.159464(6)$ (see
the black straight lines).
This is the first evidence that
the phase transition in such case
is second-order for all $\sigma$'s.
In order to confirm this query, we  also
obtained static exponents, by
performing steady numerical simulations.
In the case of a continuous transition  $\phi$ will behave  as
$\phi \sim (\alpha_c-\alpha)^{\beta}$, where  
 $\beta$ is the associated critical exponent.
In the left part of Fig. \ref{fig8}, we
show  a log-log plot of $\phi$ vs $\Delta \equiv \alpha_c-\alpha$ by using the
previous estimates for the $\alpha_c$'s. Note that both curves present slopes
 consistent with the DP value $\beta=0.276486$ (black lines), 
confirming the second-order phase transition.

We  also  studied  the dependence of
the order parameter $\phi$ on the system size $L$. At 
the criticality $\phi$ 
decays according to the power law 
 $\phi \sim L^{-\beta/\nu_{\perp}}$, where $\nu_{\perp}$ is the
critical exponent associated with the spacial length correlation. 
In fact, as showed in 
 Fig. \ref{fig8}, for  all $\alpha$'s (circles) 
$\phi$ also decays algebraically
with $L$ with critical exponents consistent with
the DP value $0.2520718$ (solid lines).
 Above conclusions remain valid  for larger $a$'s. 
For example, for $a=5$ and $\sigma=0.5$, $\phi$ 
presents an algebraic decay at  $\alpha_c=0.3318(2)$ 
with a dynamic exponent consistent with the DP one. 
 The absence of
discontinuous transition 
is understood by inspecting the density
of particles surrounded by at least one empty site ${\rho -\phi}$,
which rules the strength of the long-range interaction in such case,
as showed    
in Fig. \ref{fig8}($d$) for $\sigma=0.1$. The existence of 
infinitely absorbing states leads  ${\rho-\phi}$ to change  very mildly 
(close to the transition),
 implying that  active states with low $\phi$ are not
destabilized by the long-range interaction (even for low $\sigma$) 
and hence  no abrupt  change of $\phi$ occurs and 
the transition remains continuous.

It is worth mentioning that the absence of discontinuous transition
in not predicted by mean-field approximations. By taking  
two site correlations, the   
transition is discontinuous for $\sigma<\sigma_t$, with a  
tricritical point separating first and second lines given by
\begin{equation}
\zeta(\sigma_t) = 2\frac{1+a}{a}.
\end{equation}
Again, from this equation it follows that $\sigma_t>1$. 
The pair mean-field predicts a critical line at $\alpha_c=1/4$.

\begin{figure}
\centering
\epsfig{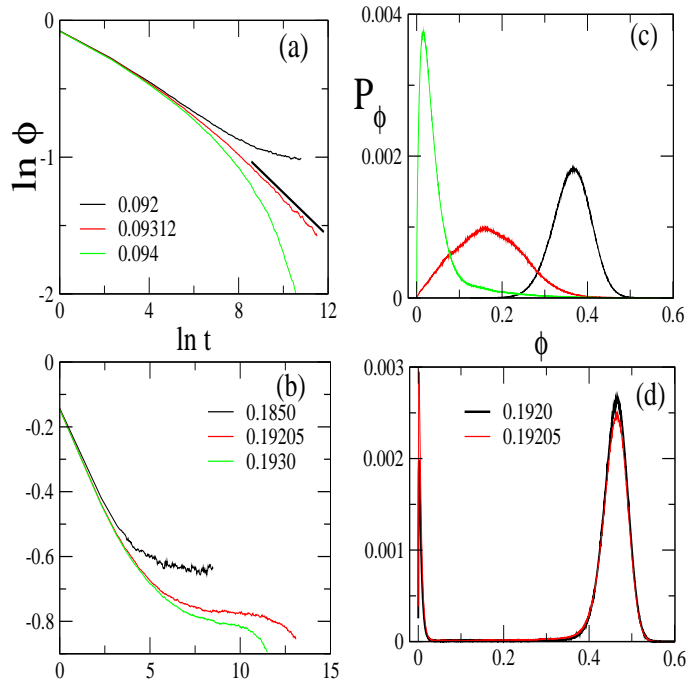}
\caption{Log-log plot of the 
time decay of $\phi$ for $\sigma=1.2$ and $\sigma=0.1$
in  $(a)$ and $(b)$, respectively.  
In $(c)$ and $(d)$, probability distribution $P_{\phi}$ vs $\phi$ for the
$\sigma-$PCP and $\sigma=1.2$  and $\sigma=0.1$. }
\label{fig9}
\end{figure}

In Fig. \ref{fig9}, we show the main results for the second
version. In $(a)$ and $(b)$ 
we compare the time decay of $\phi$ for $a=2$ with 
$\sigma=1.2$ and $\sigma=0.1$,
respectively. In the former case,  
the slope at $\alpha_c=0.09312$ agrees with the value $0.159464(6)$,
in consistency with the emergence of 
second-order transitions for $\sigma>1$. In contrast,
the decay for  $\sigma=0.1$ is slightly different from the previous case, 
suggesting a first-order transition. The phase
coexistence is confirmed by  plotting the  pair density probability 
distribution $P_\phi$,  as showed in $(c)$ and 
$(d)$ for $\sigma=1.2$ and $\sigma=0.1$, respectively.
In fact,  for $\sigma=0.1$ $P_\phi$ is bimodal. 
Similar results for $\sigma=0.5$ 
support a first-order transition for $0<\sigma<1$. 
This result  is understood  by noting that in the present case  $\phi$
 plays a similar role to $\rho$ in the  $\sigma-$CP. 
The  dynamics described by Eqs. (\ref{rate4}) and  (\ref{rate7})  
allows to relate  $\sigma-$PCP and the  $\sigma-$CP
through transformation $\eta_i\eta_{i+1}\rightarrow \eta^{'}_{i}$.
Since in the $\sigma$-CP states of low densities are disrupted
by the long range interactions, a similar
conclusion holds valid the $\sigma$-PCP. For the previous case,
such analogy can not be drown, due to the dependence with  one site occupied
in the frontier instead of two occupied sites. To close
this section we remark that    the coexistence line
also seems to move toward larger $\sigma$'s when $a$ increases, 
although the crossover between the two regimes  
is broader than the $\sigma-$CP. 

\section{Conclusion}
First-order phase transitions into absorbing states require a
robust mechanism of preventing the creation of particles in
low density regimes. 
Although there are strong evidences that they can not occur  in one-dimensional
short-range contact processes,  a long range counterpart
 by Ginelli et al. \cite{gine05} revealed such possibility.
Aimed at uncovering the fundamental mechanisms ruling
 long range interactions as an  effective mechanism
of forming compact clusters, here we  investigated thoroughly a
family of interaction rules. Our study includes 
seven contact models  grouped
in three distinct approaches. In the former, we considered weak 
long-range interactions and competition
between with frequent short-range dynamics, whereas the
latter replaces  single by cluster interactions. All
 these results  have revealed that it suffices a long-range small
 perturbation for suppressing  low density stable states,
 in the sense that the undertaken weakening long-range 
interactions are not sufficient ``strong'' for destroying compact clusters.
The mean-field approach gave us some insight to understand 
above conclusions. Except the $\sigma-$pair CP, the approximated expressions
present a general structure, predicting a changing in the phase
transition for all $a$ and  $\sigma$. However, in 
contrast with numerical results, the  critical lines present the same 
transition point $\alpha_c$ (for all $\sigma>\sigma_t$). Unlike previous cases,
the presence of infinitely many absorbing
states leads to  novel and different scenarios, depending on the 
particle structures surrounding the edge of inactive sites. 
One of them  predicts  conclusions similar to those of the
  previous models, whereas  
in the other version, the phase coexistence is  destroyed.
Such result can not be understood by the two-site mean-field theory, 
in which the phase transition also  becomes first-order for small values
of $\sigma$. 
Although the increase of fluctuations may predict
a second-order  transition for small $\sigma$, 
we believe that, in the present case, 
a very large order of approximation would be required
to reproduce a continuous transition. 
In summary,  long-range interactions constitute an effective dynamics
to provide a discontinuous phase transition, even for 
extreme  limits undertaken here.
As a final remark, we should mention that
the effect of other dynamics 
such as diffusion and quenched disorder
and its competition with long-range interactions should be 
investigated in a further contribution.

\section*{ACKNOWLEDGEMENT}
We acknowledge Miguel A. Mu\~noz for a critical
reading of this manuscript and useful suggestions.  
The financial support from   CNPQ is also acknowledged.

\end{document}